\documentclass[10pt,conference]{IEEEtran}

\usepackage{amssymb}
\usepackage{amsmath}
\usepackage{cite}
\usepackage{url}
\usepackage{xcolor}
\usepackage{cite,graphicx,amsmath,amssymb}
\usepackage{fancyhdr}
\usepackage{mdwmath}
\usepackage{mdwtab}
\usepackage{caption}
\usepackage{amsthm}
\usepackage{setspace}
\usepackage{xcolor}
\usepackage{algorithm}
\usepackage{algorithmic}
\usepackage{subfig}
\usepackage{multirow}
\usepackage{makecell}
\usepackage{mathtools}

\graphicspath{{./src/img/}}

\newtheorem{remark}{Remark}
\newtheorem{theorem}{Theorem}

\newtheorem{lemma}{Lemma}

\newtheorem{corollary}{Corollary}

\allowdisplaybreaks
\title{Active-Sensing-Based Beam Alignment for Near Field MIMO Communications}

\author{
    \IEEEauthorblockN{Hao Jiang\IEEEauthorrefmark{1}, Zhaolin Wang\IEEEauthorrefmark{1}, and Yuanwei Liu\IEEEauthorrefmark{1}\\
\IEEEauthorblockA{\IEEEauthorrefmark{1}Queen Mary University of London, London, UK.
\\E-mail: \{hao.jiang, zhaolin.wang, yuanwei.liu\}@qmul.ac.uk}}
}

\begin{document}
\maketitle
\begin{abstract}
An active-sensing-based learning algorithm is proposed to solve the near-field beam alignment problem with the aid of wavenumber-domain transform matrices (WTMs).
Specifically, WTMs can transform the antenna-domain channel into a sparse representation in the wavenumber domain.
The dimensions of WTMs can be further reduced by exploiting the dominance of line-of-sight (LoS) links.
By employing these lower-dimensional WTMs as mapping functions, the active-sensing-based algorithm is executed in the wavenumber domain, resulting in an acceleration of convergence.
Compared with the codebook-based beam alignment methods, the proposed method finds the optimal beam pair in a ping-pong fashion, thus avoiding high training overheads caused by beam sweeping.
Finally, the numerical results validate the effectiveness of the proposed method.    
\end{abstract}

\section{Introduction}
During the past few decades, the world has witnessed the great convenience brought by the exponential growth of wireless connectivity.
As a double-edged sword, the formidable spectrum demands brought by such massive connections pose an unprecedented challenge to the next-generation wireless communication technologies.
As a remedy, the high-frequency bands, such as millimeter wave (mmWave) and terahertz (THz) bands, can potentially provide an enormous bandwidth with an order of tens up to a hundred gigahertz (GHz), for supporting massive connections \cite{Shafie2023Therahertz, wang2018millimeter}.

However, communications over such high frequency bands inevitably suffers from severe propagation losses, which can lead to degradation of signal strength \cite{qurratulain2023machine, Giordani2019tutorial}.
To build reliable communication links, the beam management framework is adopted by 5G new radio (NR) for multiple-input multiple-output (MIMO) communication networks\cite{heng2023grid, qurratulain2023machine}, in which beam alignment is a vital step.
During the beam alignment process, transmitters and receivers need to sweep their respective predefined codebooks, i.e., grid-of-beams, in an exhaustive manner \cite{zhang2020beam} or a hierarchical manner \cite{xiao2026hierarchical}, to find the best directional beam pair.
In these methods, there is a balance between the codebook size and the resolution of beam alignment.
Specifically, finer codewords, i.e., narrower beams, will incur a larger codebook size but a higher resolution of beam alignment.
To strike a good balance between them, the authors of \cite{heng2023grid} proposed a grid-free beam alignment approach, in which a machine learning technique was used to find site-specific optimal codebooks. 
Nevertheless, these methods can still cause significant overheads, due to the exhaustive beam sweeping.
To address this issue, the authors of \cite{jiang2023active} proposed an active-sensing approach to carry out two-side beam alignment using ping-pong pilots, which excepted the usage of codebooks and therefore reduced overheads.

The above discussion is limited in the scope of far-field communication (FFC) scenarios.
In the next-generation wireless systems, the prevailing trend of deploying extremely-large (XL) MIMO arrays extends the near-field communication (NFC) region to tens of or even hundreds of meters.
This causes that the planar wavefront assumption in FFC fail to capture the characteristics of wireless channels \cite{liu2023near}.
In contrast to FFC, the spherical wavefront needs to be considered in NFC, which is determined not only by the angular domain but also by the distance domain.
In this case, the codebook-based beam alignment methods entails sampling both of the domains to construct NFC codebooks\cite{zhang2023beam, chen2023hierarchical, wu0two, zhang2022fast}, resulting in unacceptable overheads and intolerable delays.
The scenario becomes even worse in MIMO systems, since beam sweeping must be carried out on both sides.
On the contrary, active sensing is a promising approach to address this issue, which exploits alternating ping-pong pilots to align beams at the MIMO transceivers, thus avoiding the need for sweeping over high-dimensional codebooks \cite{jiang2023active}. 

Against the above discussion, this paper proposes an active-sensing-based learning algorithm for NFC MIMO systems, without requiring high-dimensional codebooks.
This algorithm exploits simple channel representations in the wavenumber domain to accelerate the beam alignment process and cut down the computational complexity.
Specifically, by harnessing the dominance of the line-of-sight (LoS) path in high-frequency bands \cite{tang2023line}, the antenna-domain channel representation can be truncated into a simple diagonal and low-dimensional structure using wavenumber-domain transform matrices.
Therefore, by using these matrices as mapping functions, the active-sensing-based algorithm can align the beams according to simple wavenumber-domain channel representations. 
Finally, numerical results demonstrate the fast convergence and near-optimal performance of the proposed algorithm.

\section{System Model}
\begin{figure}[h!]
	\centering
	\includegraphics[width=0.85\linewidth]{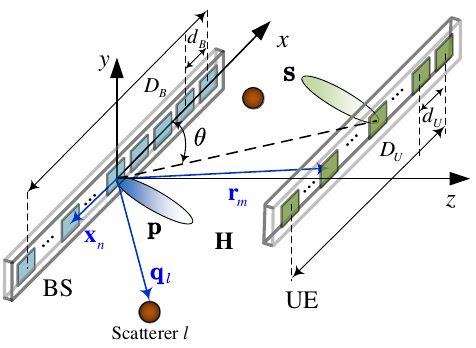}
	\caption{An illustration of a NFC MIMO system.}
	\label{fig:system_model}
\end{figure}
As depicted in Fig. 1, we consider a narrowband NF MIMO communication system, which consists of a base station (BS) equipped with an $N$-antenna uniform linear array (ULA) with $N = 2 \tilde{N} - 1$ and an user equipment (UE) with an $M$-antenna ULA with $M = 2 \tilde{M} - 1$.
The ULAs are assumed to locate on the $xz$-plane and parallel to one other.
The angle between the center of the ULA at the UE and the $x$-axis is denoted by $\theta$. 
The array apertures of the ULAs at the BS and the UE can be calculated by $D_{\text{B}} = (N - 1)d_{\text{B}}$ and $D_{\text{U}} = (M-1)d_{\text{U}}$, respectively, with $d_{\text{B}}$ and $d_{\text{U}}$ denoting the antenna spacing.
The distance between the center points of the BS and the UE are assumed to be shorter than than Rayleigh distance, i.e., $d_{\text{BU}} < \frac{2(D_{\text{B}} + D_{\text{U}})^2}{\lambda}$, where $\lambda$ denotes the signal wavelength.

\subsection{Near-field Channel Model}
To capture the near-field characteristics, we adopt the non-uniform spherical-wave (NUSW) channel model \cite{liu2023near} to represent the MIMO channel, which consists of a LoS channel and $L$ NLoS channels caused by randomly deployed scatterers.
Let ${\bf x}_n = [x_n^{(x)}, y_n^{(x)},  z_n^{(x)}]^T \in \mathbb{R}^{3 \times 1}$, where $n \in \{-\tilde{N}, ..., \tilde{N}\}$, denote the coordinate of the $n$-th antenna at the BS, ${\bf r}_m = [x_m^{(r)}, y_m^{(r)},  z_m^{(r)}]^T \in \mathbb{R}^{3 \times 1}$, where $ m \in\{-\tilde{M}, ..., \tilde{M}\}$ denote the $m$-th antenna at the UE, and $\mathbf{q}_l=[ x_{l}^{(q)},y_{l}^{(q)},z_{l}^{(q)} ]^T \in \mathbb{R}^{3 \times 1}$ denote the coordinate of the $l$-th scatterer.
Based on the NUSW model, the $(m,n)$-th entry of the LoS channel matrix between the BS and the UE is characterized by
\begin{align}
	\left[ \mathbf{H}_{\mathrm{LoS}} \right] _{m,n}=\frac{\lambda}{4 \pi \left\| \mathbf{r}_m-\mathbf{x}_n \right\|}e^{-jk_0\left\| \mathbf{r}_m-\mathbf{x}_n \right\|},
\end{align} 
where $k_0\triangleq 2\pi /\lambda$ denotes the wavenumber.
On the contrary, the NLoS channel can be written as a combination of two multiple-input single-output (MISO) channels, which can be modeled by
\begin{align}
	\mathbf{H}_{\mathrm{NLoS}}=\sum_{l=1}^L{\beta _l\mathbf{a}_{\mathrm{B}}\left( \mathbf{q}_l \right) \mathbf{a}_{\mathrm{U}}^{T}\left( \mathbf{q}_l \right)},
	,
\end{align}
where $\beta _l$ denotes the path-loss and random reflection coefficient that is drawn from $\mathcal{CN}(0, {\sigma_l}^2)$. 
Vectors $\mathbf{a}_{\mathrm{B}}\left( \mathbf{q}_l \right) \in \mathbb{C}^{N \times 1}$ and $\mathbf{a}_{\mathrm{U}}\left( \mathbf{q}_l \right) \in \mathbb{C}^{M \times 1}$ denote the array response vectors at the BS and the UE, respectively, which can be expressed as follows:
\begin{align}
	\mathbf{a}_{\mathrm{B}}\left( \mathbf{q}_l \right) &=\left[ e^{-jk_0\left\| \mathbf{q}_l-\mathbf{x}_{-\tilde{N}} \right\|},...,e^{-jk_0\left\| \mathbf{q}_l-\mathbf{x}_{\tilde{N}} \right\|} \right] ^T, \\
	\mathbf{a}_{\mathrm{U}}\left( \mathbf{q}_l \right) &=\left[ e^{-jk_0\left\| \mathbf{q}_l-\mathbf{r}_{-\tilde{M}} \right\|},...,e^{-jk_0\left\| \mathbf{q}_l-\mathbf{r}_{\tilde{M}} \right\|} \right] ^T.
\end{align} 
By considering both of the above, the NFC channel can be written as
\begin{align}
	\mathbf{H}=\mathbf{H}_{\mathrm{LoS}} + \mathbf{H}_{\mathrm{NLoS}}.
\end{align}
It is noted that in high-frequency bands, communication channels are LoS-dominated and NLoS-assisted due to severe scattering losses.

\subsection{Signal Model}
To enhance the energy efficiency, the analog beamforming architecture is assumed for both the BS and the UE, where the antenna array is connected to a single radio-frequency (RF) chain through a large number of phase shifters (PSs). 
Let $\mathbf{p} \in \mathbb{C}^{N \times 1}$ and $\mathbf{s} \in \mathbb{C}^{M \times 1}$ denote the the probing and the sensing beam realized by the PSs at the BS and the UE, respectively. 
Due to the hardware constraint of PSs, vectors $\mathbf{p}$ and $\mathbf{s}$ are subject to a constant-modulus constraint, i.e., $| \left[ \mathbf{p} \right] _{n} |=\frac{1}{\sqrt{N}}$ and $ | \left[ \mathbf{s} \right] _{m} |=\frac{1}{\sqrt{M}}$.
Assuming that the reciprocity of $\mathbf{H}$ holds, the received signal at the UE via downlink transmission and that at the BS by uplink transmission are given by 
\begin{align}
	{\mathbf{y}}^{\mathrm{U}}&=\mathbf{Hp}c_{\mathrm{B}}+{\bf n}_{\mathrm{B}}, \\
	{\mathbf{y}}^{\mathrm{B}}&=\mathbf{H}^T\mathbf{s}c_{\mathrm{U}}+{\bf n}_{\mathrm{U}},
\end{align}
where $c_{\text{B}} \in \mathbb{C}$ and  $c_{\text{U}} \in \mathbb{C}$ denote the baseband pilot signals at the BS and the UE, respectively, satisfying $|c_\text{B}|^2 = P_\text{B}$ and $|c_\text{U}|^2 = P_\text{U}$, and 
$\mathbf{n}_{\mathrm{B}} \sim \mathcal{CN}(0, \sigma_{\mathrm{B}}^2 {\mathbf{I}}_N)$ and $\mathbf{n}_{\mathrm{U}} \sim \mathcal{CN}(0, \sigma_{\mathrm{U}}^2{\mathbf{I}}_M )$ denote the circularly symmetric complex Gaussian noise (CSCG) with power $\sigma_{\mathrm{B}}^2$ and  $\sigma_{\mathrm{U}}^2$, respectively.  
\section{Problem Formulation}
The objective of beam alignment is to maximize the beam gain, which is defined as $U\left( \mathbf{s},\mathbf{p} \right) =| \mathbf{s}^T\mathbf{Hp} | ^2=| \mathbf{p}^T\mathbf{H}^T\mathbf{s} | ^2$, without the knowledge of channel state information (CSI).
The NFC beam alignment problem can be formulated as follows: 
\vspace{-0.2cm}
\begin{subequations}
	\begin{align} \label{object}
		& \hspace{-1cm} \max_{\mathbf{s},\mathbf{p}}~~~U\left( \mathbf{s},\mathbf{p} \right) \tag{8}\\ \label{c-1}
		\mathrm{s.t.} \quad &\left| \left[ \mathbf{p} \right] _n \right|=\frac{1}{\sqrt{N}}, ~~\forall n \in \{1, ..., N\} \\
		\label{c-2}
		& \left| \left[ \mathbf{s} \right] _m \right|=\frac{1}{\sqrt{M}}, ~~\forall m \in \{1, ..., M\} 
	\end{align}
\end{subequations}
The above problem is generally challenging to solve due to the non-convexity caused by \eqref{c-1} and \eqref{c-2}, and the high dimensions of $\bf s$ and $\bf p$ caused by the density of antennas.
To address these challenges, in the following section, we proposed an active-sensing-based method with the aid of wavenumber-domain transform matrices, which can solve \eqref{object} in the absence of feedback links between the transceivers.
\section{Proposed Solution}
In this section, the solution to \eqref{object} is proposed, which includes two parts.
As a preliminary preparation, in the first part, the Fourier plane-wave series expansion is introduced, using which the sparse NFC MIMO channel structure can be revealed. 
In the other part, jointly using the Fourier plane-wave series expansion and the deep learning technique, an active-sensing-based algorithm is proposed. 

\subsection{Wavenumber-Domain Representation of Channels}
The basic idea of wavenumber-domain representations of NFC channels is that a spherical wavefront can be approximated by a superposition of multiple planar wavefronts.  
According to the methodology in \cite{pizzo2021fourier,tang2023line}, the spatial impulse response $h_{m,n} = \left[ \mathbf{H} \right] _{m,n}$ can be derived based on a 4D Fourier plane-wave representation, which is given by
\begin{align}
	&h_{m,n}=\frac{1}{\left( 2\pi \right) ^2}\iiiint\limits_{\mathcal{D} _{\boldsymbol{\kappa }}\times \mathcal{D} _{\mathbf{k}}}^{}{a_{\mathrm{U}}\left( \boldsymbol{\kappa },\mathbf{r}_m \right) H_{\mathrm{a}}\left( x^{\left( \kappa \right)},y^{\left( \kappa \right)},x^{\left( k \right)},y^{\left( k \right)} \right)} \notag \\
	&\qquad \qquad \qquad \qquad \times {a_{\mathrm{B}}\left( \mathbf{k},\mathbf{x}_n \right) dx^{\left( \kappa \right)}dy^{\left( \kappa \right)}dx^{\left( k \right)}dy^{\left( k \right)}}, \label{eq:ft_cont}
\end{align}
where $H_{\mathrm{a}}( x^{\left( \kappa \right)},y^{\left( \kappa \right)},x^{\left( k \right)},y^{\left( k \right)} )$ denotes the coupling coefficient between the transmitted and received plane waves, that are respectively given by $\boldsymbol{k}=[ x^{\left( k \right)},y^{\left( k \right)},\gamma \left( x^{\left( k \right)},y^{\left( k \right)} \right) ] ^T$ and $\boldsymbol{\kappa }=[ x^{\left( \kappa \right)},y^{\left( \kappa \right)},\gamma \left( x^{\left( \kappa \right)},y^{\left( \kappa \right)} \right) ] ^T$, where $\gamma \left( a,b \right) \triangleq \sqrt{k_0^2-a^2-b^2}$;
the transmit and receive responses at $\mathbf{x}_m$ and $\mathbf{r}_m$ can be respectively expressed as
$a_{\mathrm{B}}\left( \mathbf{k},\mathbf{x}_n \right) =e^{-j\mathbf{k }^T\mathbf{x}_n}$ and $a_{\mathrm{U}}\left( \boldsymbol{\kappa },\mathbf{r}_m \right) =e^{j\boldsymbol{\kappa }^T\mathbf{r}_m}$.
By excluding the reactive near-field region, which is typically few wavelengths, $\gamma ( x^{\left( k \right)},y^{\left( k \right)} ) $ and $\gamma ( x^{( \kappa )},y^{\left( \kappa \right)} ) $ are real-valued.
Then, the discrete sets $\mathcal{D}_{\mathbf{k}}$ and $\mathcal{D} _{\boldsymbol{\kappa }}$ can be specified by
\begin{align}
	\mathcal{D} _{\mathbf{k}} &\triangleq \left\{ ( x^{\left( k \right)},y^{\left( k \right)} ) \in \mathbb{R} ^2:\left( x^{\left( k \right)} \right) ^2+\left( y^{\left( k \right)} \right) ^2\leqslant k_0^2 \right\}, \label{eq:D_k}
	\\
	\mathcal{D} _{\boldsymbol{\kappa }} &\triangleq \left\{ ( x^{\left( \kappa \right)},y^{\left( \kappa \right)} ) \in \mathbb{R} ^2:\left( x^{\left( \kappa \right)} \right) ^2+\left( y^{\left( \kappa \right)} \right) ^2\leqslant k_0^2  \right\}. \label{eq:D_kap}
\end{align}
According to \cite[Theorem 2]{pizzo2021fourier}, when antenna arrays are electronmagnetically large, i.e., their apertures compares with $\lambda$, MIMO channel can be approximated by a finite number of plane waves.
In this paper, since the ULAs are assumed to be deployed on the $xz$-plane, we have $y_{m}^{(r)}=y_{n}^{(x)}=0$.
Then, the discrete version of \eqref{eq:D_k} and \eqref{eq:D_kap} can be written as $\mathcal{G} _{\boldsymbol{\kappa }} \triangleq \{ i\in \mathbb{Z} : (\frac{\lambda i}{D_{\mathrm{U}}})^2  \leqslant 1 \}$ and $\mathcal{G} _{\mathbf{k}}\triangleq \{ j\in \mathbb{Z} : (\frac{\lambda j}{D_{\mathrm{B}}})^2 \leqslant 1 \}$, respectively.
With these discrete sets, equation \eqref{eq:ft_cont} can be discretized to   
\begin{align}
	&h_{m,n}  \approx \notag \\
	&~~\sum_{i\in \mathcal{G} _{\boldsymbol{\kappa }},j\in \mathcal{G} _{\mathbf{k}}}{\phi _{\mathrm{U}}\left( i,x_{m}^{(r)} \right) \tilde{H}_{\mathrm{a}}\left( i,j,z_{m}^{(r)},z_{n}^{(x)} \right) \phi _{\mathrm{B}}^{*}\left( j,x_{n}^{(x)} \right)}
	, \label{eq:H_discrete} 
\end{align}
where
\begin{align}
	\tilde{H}_{\mathrm{a}}\left( i,j,z_{m}^{(r)},z_{m}^{(x)} \right) &= e^{j\gamma \left( \frac{2\pi i}{D_{\mathrm{U}}} \right) z_{m}^{(r)}} {H}_{\mathrm{a}}\left( i,j \right) e^{-j\gamma \left( \frac{2\pi j}{D_{\mathrm{B}}} \right) z_{n}^{(x)}}, \\
	\phi_{\mathrm{U}} \left( i,x_{m}^{(r)} \right) &=e^{j\left( \small{\frac{2\pi i}{D_{\mathrm{U}}}x_{m}^{(r)}} \right)},
	\\
	\phi_{\mathrm{B}} \left( j,x_{n}^{(x)} \right) &=e^{j\left( \small{\frac{2\pi j}{D_{\mathrm{B}}}x_{n}^{(x)}} \right)}.
\end{align} 
More compactly, equation \eqref{eq:H_discrete} can be described by  
\begin{align}
	\mathbf{H} \approx \sqrt{MN} \mathbf{\Phi }_{\mathrm{U}}\tilde{\mathbf{H}}_{\mathrm{a}}\mathbf{\Phi }_{\mathrm{B}}^{H},
\end{align}
where $\mathbf{\Phi }_{\mathrm{U}}$ and $\mathbf{\Phi }_{\mathrm{B}}$ denote the semi-unitary wavenumber-domain transform matrices and are respectively given by 
\begin{align}
	\mathbf{\Phi }_{\mathrm{U}}&=\left[ ...,\boldsymbol{\phi }_{\mathrm{U},i}^{},... \right] ^T\in \mathbb{C} ^{M\times \left| \mathcal{G} _{\boldsymbol{\kappa }} \right|},
	\\
	\mathbf{\Phi }_{\mathrm{B}}&=\left[ ...,\boldsymbol{\phi }_{\mathrm{B},j}^{},... \right] ^T\in \mathbb{C} ^{N\times \left| \mathcal{G} _{\mathbf{k}} \right|},
\end{align}
where $i\in \mathcal{G} _{\boldsymbol{\kappa }}^{}$, $j\in \mathcal{G} _{\mathbf{k}}^{}$, $[ \tilde{\mathbf{H}}_{\mathrm{a}} ] _{i,j}=\tilde{H}_{\mathrm{a}}(i,j,z_{m}^{(r)},z_{n}^{(x)})$, $\boldsymbol{\phi }_{\mathrm{U},i}^{}=\frac{1}{\sqrt{M}}[ \phi _{\mathrm{U}}^{}( i,x_{1}^{(r)} ) ,...,\phi _{\mathrm{U}}^{}( i,x_{M}^{(r)} ) ] ^T\in \mathbb{C} ^{M\times 1}
$, and $\boldsymbol{\phi }_{\mathrm{B},j}^{}=\frac{1}{\sqrt{N}}[ \phi _{\mathrm{B}}^{}( i,x_{1}^{(\mathrm{x)}} ) ,...,\phi _{\mathrm{B}}^{}( i,x_{N}^{(\mathrm{x)}} )] ^T\in \mathbb{C} ^{N\times 1}
$.
According to \cite{pizzo2021fourier}, $\tilde{\mathbf{H}}_{\mathrm{a}} $  is semi-unitary equivalent to ${\mathbf{H}}$, meaning that they have the identical the most significant singular values.
Moreover, compared with ${\mathbf{H}}$, $\tilde{\mathbf{H}}_{\mathrm{a}}$ has a diagonal and sparse structure.
By using the definition of semi-unitary, we have $\tilde{\mathbf{H}}_{\mathrm{a}}\propto\mathbf{\Phi }_{\mathrm{U}}^{H}\mathbf{H\Phi }_{\mathrm{B}}^{}$.
\begin{figure} 
	\centering
	\subfloat[$|\mathbf{H}|$]{
		\includegraphics[width=0.5\linewidth]{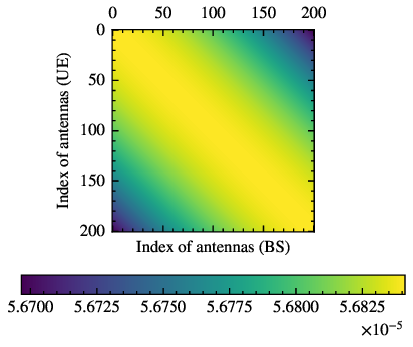}}
	\subfloat[$|\tilde{\mathbf{H}}_{\mathrm{a}}|$]{
		\includegraphics[width=0.5\linewidth]{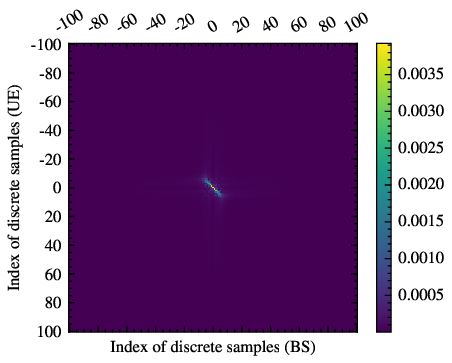}}
	\label{1b}\\
	\subfloat[$|\tilde{\mathbf{H}}_{\mathrm{e}}|$]{
		\includegraphics[width=0.5\linewidth]{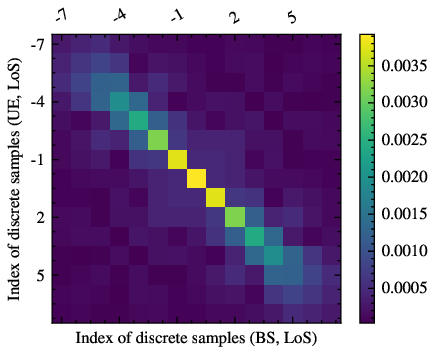}}
	\label{1c}
	\caption{The illustration of channel representations of a LoS NFC MIMO channel. Fig. (a), (b), and (c) are the channel representations in the antenna domain, the wavenumber domain, and the truncated wavenumber domain, respectively.} 
	\label{fig:waveform_domain} 
\end{figure}

For XL-MIMO communication systems, the propagation is dominated by LoS channels. 
Therefore, it is reasonable to describe the channel using a low-dimensional sub-space of the wavenumber domain, composed by LoS paths, while omitting NLoS paths. 
\begin{figure*}[ht!]
	\centering
	\includegraphics[width=0.9\linewidth]{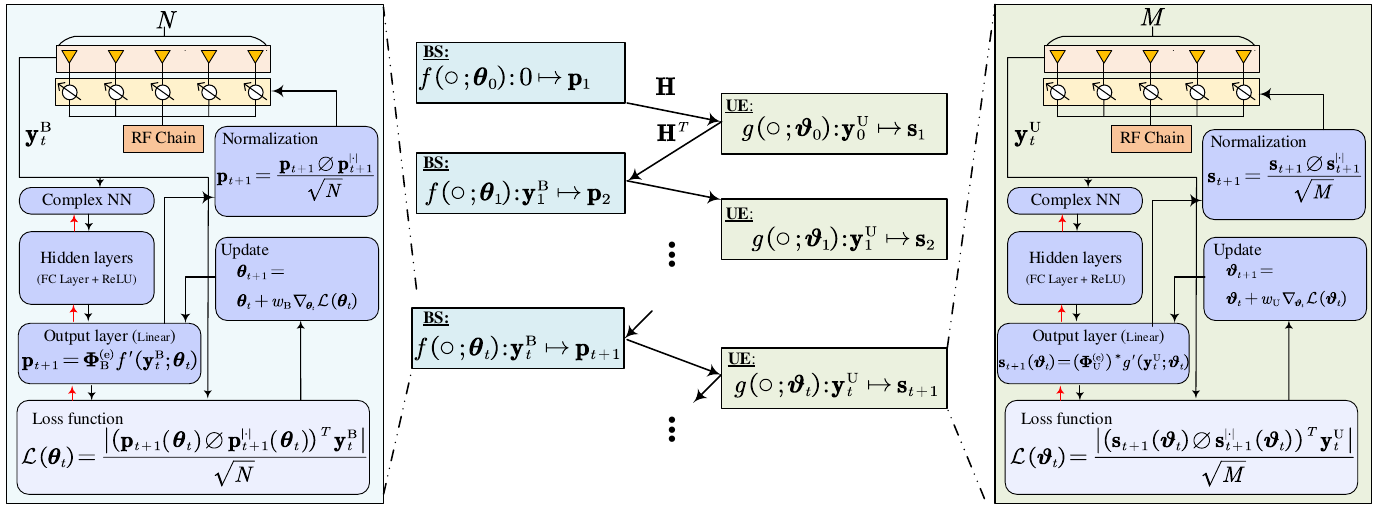}
	\caption{An overview of proposed active-sensing-based beam alignment algorithm. Data/gradient flows is denoted by the black/red line.}
	\label{fig:NN}
\end{figure*}
Therefore, to extract the LoS components, the boundaries of the LoS sub-space needs to be defined. 
According to \cite{tang2023line}, for the paralleled ULAs case, the boundaries of the LoS sub-space at point $\mathbf{r}$, generated by the source at $\bf s$,  can be defined by
\begin{align}
	w\left( \mathbf{r} \right) &=k_{\max}^{\left( \text{e} \right)}-k_{\min}^{\left( \text{e} \right)} \notag 
	\\&=k_0\max_{\mathbf{s}} \hat{\mathbf{r}}^T\left( \mathbf{r},\mathbf{s} \right) \hat{\mathbf{x}}-k_0\min_{\mathbf{s}} \hat{\mathbf{r}}^T\left( \mathbf{r},\mathbf{s} \right) \hat{\mathbf{x}}
	,
\end{align} 
where $\hat{\mathbf{r}}\left( \mathbf{r},\mathbf{s} \right) =\left( \mathbf{r}-\mathbf{s} \right) /\left\| \mathbf{r}-\mathbf{s} \right\| \in \mathbb{R} ^{3\times 1}$ denotes the normalized directional vector from $\bf s$ to $\bf r$ and $\hat{\mathbf{x}}=\left[ 1,0,0 \right] ^T\in \mathbb{R} ^{3\times 1}$.
Based on the above boundaries, the truncated discrete versions of \eqref{eq:D_k} and \eqref{eq:D_kap} can be expressed as
\begin{align}
	\mathcal{G} _{\mathbf{k}}^{\left( \mathrm{e} \right)} &= \{j\in \mathbb{Z} :k_{\min}^{\left( \mathrm{e} \right)}\leqslant  \frac{2\pi j}{D_{\mathrm{B}}} \leqslant k_{\max}^{\left( \mathrm{e} \right)}\}
	, \\
	\mathcal{G} _{\boldsymbol{\kappa }}^{\left( \mathrm{e} \right)} &=\{i\in \mathbb{Z} :k_{\min}^{\left( \mathrm{e} \right)}\leqslant  \frac{2\pi i}{D_{\mathrm{U}}} \leqslant k_{\max}^{\left( \mathrm{e} \right)}\}
	.
\end{align}
With the above truncated sets, we can further truncate $\mathbf{\Phi }_{\mathrm{U}}$ and $\mathbf{\Phi }_{\mathrm{B}}$ to $\mathbf{\Phi }_{\mathrm{U}}^{\left( \mathrm{e} \right)}$ and $\mathbf{\Phi }_{\mathrm{B}}^{\left( \mathrm{e} \right)}$, which are given by 
\begin{align}
	\mathbf{\Phi }_{\mathrm{U}}^{\left( \mathrm{e} \right)}&=\left[ ...,\boldsymbol{\phi }_{\mathrm{U},i}^{},... \right]^T \in \mathbb{C} ^{M\times \left| \mathcal{G} _{\boldsymbol{\kappa }}^{\left( \mathrm{e} \right)} \right|}, \label{eq:trans_U_eq}
	\\
	\mathbf{\Phi }_{\mathrm{B}}^{\left( \mathrm{e} \right)}&=\left[ ...,\boldsymbol{\phi }_{\mathrm{B},j}^{},... \right]^T \in \mathbb{C} ^{N\times \left| \mathcal{G} _{\mathbf{k}}^{\left( \mathrm{e} \right)} \right|}. \label{eq:trans_B_eq}
\end{align} 
where $i \in \mathcal{G} _{\boldsymbol{\kappa }}^{\left( \mathrm{e} \right)}$ and $j \in \mathcal{G} _{\mathbf{k}}^{\left( \mathrm{e} \right)}$.
It is noted that compared with the original wavenumber-domain transform matrices, the truncated ones have lower dimensions, i.e., $|\mathcal{G} _{\mathbf{k}}^{\left( \mathrm{e} \right)} | < |\mathcal{G} _{\mathbf{k}} |$ and  $|\mathcal{G} _{\boldsymbol{\kappa }}^{\left( \mathrm{e} \right)} | < |\mathcal{G} _{\boldsymbol{\kappa }}|$. 
Using $\mathbf{\Phi }_{\mathrm{U}}^{\left( \mathrm{e} \right)}$ and $\mathbf{\Phi }_{\mathrm{B}}^{\left( \mathrm{e} \right)}$, the channel can be expressed by
\begin{align}
	\mathbf{H} \approx \sqrt{MN}\mathbf{\Phi }_{\mathrm{U}}^{\left( \mathrm{e} \right)}\tilde{\mathbf{H}}_{\mathrm{e}}\left( \mathbf{\Phi }_{\mathrm{B}}^{\left( \mathrm{e} \right)} \right) ^H. \label{eq:effective}
\end{align}
Similarly, we have $\tilde{\mathbf{H}}_{\mathrm{e}}\propto( \mathbf{\Phi }_{\mathrm{U}}^{\left( \mathrm{e} \right)} ) ^H\mathbf{H\Phi }_{\mathrm{B}}^{\left( \mathrm{e} \right)}$.

In Fig. \ref{fig:waveform_domain}, the channel representations are provided.
The numbers of antennas are set to $M=N=201$, the carrier frequency is set to $28$ GHz, and
$d_{\text{BU}}$ and $\theta$ are set to $15~m$ and $90^\circ$.
The original NFC channel $\mathbf{H}$ is converted to a sparse structure, i.e., $\tilde{\mathbf{H}}_{\mathrm{a}}$, where channel power is concentrated in a small sub-space on its diagonal.
Further, via the truncated transform matrices, a low-dimensional sub-space denoted by $\tilde{\mathbf{H}}_{\mathrm{e}}$ can be extracted.

\subsection{Active-Sensing-Based Learning Algorithm}
Based on the above, by using the semi-unitary property of the transform matrices, the beam alignment can be carrier out in the wavenumber domain, where the channel representations are simple. 
Specifically, letting ${\bf p}^\prime \in \mathbb{C}^{| \mathcal{G} _{\boldsymbol{k }}^{\left( \mathrm{e} \right)} | \times 1}$ and ${\bf s}^\prime \in \mathbb{C}^{| \mathcal{G} _{\boldsymbol{\kappa }}^{\left( \mathrm{e} \right)} | \times 1}$ be the probing and sensing beams in the wavenumber domain, they can be mapped to the antenna domain via $\mathbf{p}=\mathbf{\Phi }_{\mathrm{B}}^{\left( \mathrm{e} \right)}\mathbf{p}^{\prime}
$ and $\mathbf{s}=( \mathbf{\Phi }_{\mathrm{U}}^{\left( \mathrm{e} \right)} ) ^*\mathbf{s}^{\prime}$, respectively.
Therefore, we have 
\begin{align}
	&U\left( \mathbf{s},\mathbf{p} \right) =| \mathbf{s}^T\mathbf{Hp} | ^2 \overset{\text{(a)}}{=}| \left( \mathbf{s}^{\prime} \right) ^T\tilde{\mathbf{H}}_{\mathrm{e}}\mathbf{p}^{\prime} | ^2,
\end{align}
where $\text{(a)}$ can be derived by substituting \eqref{eq:effective} into the expression.
In this case, low-dimensional ${\bf p}^\prime$ and ${\bf s}^\prime$ can be optimized according to $\tilde{\mathbf{H}}_{\mathrm{e}}$, which can simplify the beam alignment problem.  
Motivated by this idea, we propose an active-sensing approach to solve \eqref{object}, which is aided by the truncated wavenumber-domain transform matrices.
The proposed algorithm can be divided into two parts, i.e., a BS NN module and a UE NN module.
These two modules optimize their own beamforming vectors to maximize $U\left( \mathbf{s},\mathbf{p} \right)$, in a ping-pong fashion, based on their local observations.

\subsubsection{BS NN Module}
The BS NN module is a deep neural network parametrized by $\boldsymbol{\theta}$, which is a vector composed by trainable parameters in layers.
Its function is to map the received signal at time $t~(t>0)$, i.e., $\mathbf{y}_t^{\text{B}} \in \mathbb{C}^{N \times 1}$, to a probing beam $\mathbf{p}_{t+1}\in \mathbb{C}^{N \times 1}$ at time $t+1$, which can be described by $f\left( \circ ;\boldsymbol{\theta }_t \right) :\mathbf{y}_t^{\text{B}} \mapsto \mathbf{p}_{t+1}$.
Particularly, at the initial stage, i.e., $t=0$, the BS NN module can be fed with any suitable vectors to initialize the ping-pong process.
In our case, a zero vector is fed as the initial input.
The whole module can be seen as a product of two sub-modules, i.e., a complex NN mapping function and a wavenumber-domain mapping function, and can be described by
\begin{align}
	\mathbf{p}_{t+1}(\boldsymbol{\theta }_t)=\mathbf{\Phi }_{\mathrm{B}}^{\left( \mathrm{e} \right)}f^{\prime}\left( \mathbf{y}_t^{\text{B}};\boldsymbol{\theta }_t \right), \label{eq:bs_mapping}
\end{align}
where $f^{\prime}\left( \circ;\boldsymbol{\theta }_t \right): \mathbb{C}^{N\times1} \mapsto \mathbb{C}^{| \mathcal{G} _{\boldsymbol{k }}^{\left( \mathrm{e} \right)} | \times 1}$ denotes the complex NN mapping function, which maps the received signal to a wavenumber-domain vector, and $
\mathbf{\Phi }_{\mathrm{B}}^{\left( \mathrm{e} \right)}: {\mathbb{C}}^{|{\mathcal{G} _{k}^{\left( \mathrm{e} \right)}} | \times 1} \longmapsto {\mathbb{C}}^{N\times 1}$ then maps the wavenumber-domain vector to an antenna-domain probing vector. 
Furthermore, to satisfy \eqref{c-1}, vector $\mathbf{p}_{t+1}(\boldsymbol{\theta }_t)$ is normalized according to $\mathbf{p}_{t+1}=\frac{1}{\sqrt{N}}\mathbf{p}_{t+1}\oslash \mathbf{p}_{t+1}^{\left| \cdot \right|}$, where $\oslash$ is the element-wise division and $\mathbf{a}_{}^{\left| \cdot \right|}$ is the element-wise magnitude of vector $\bf a$.
Based on \eqref{object}, the optimal $\boldsymbol{\theta }$ can maximize the received signal power, i.e., $U\left( \mathbf{s},\mathbf{p} \right)$, for a given sensing beam $\mathbf{s}$.
Hence, the loss function is formulated as 
\begin{align}
	\mathcal{L} \left( \boldsymbol{\theta }_t \right) =\frac{1}{\sqrt{N}}\left| \left( \mathbf{p}_{t+1}^{}(\boldsymbol{\theta }_t)\oslash \mathbf{p}_{t+1}^{\left| \cdot \right|}(\boldsymbol{\theta }_t) \right) ^T\mathbf{y}_{t}^{\mathrm{B}} \right|.
	 \label{loss-bs}
\end{align}
The update rule for $\boldsymbol{\theta }$ can be expressed as $\boldsymbol{\theta }_{t+1}=\boldsymbol{\theta }_t+w_{\mathrm{B}}\nabla _{\boldsymbol{\theta }_t}\mathcal{L} \left( \boldsymbol{\theta }_t \right)$, where $w_{\mathrm{B}}$ is the learning rate on the BS side.
It is noted that since we are maximizing the loss function, gradient ascent is utilized here.

\subsubsection{UE NN Module}
Similar to the BS NN module, the UE NN module is parameterized by $\boldsymbol{\vartheta}$. Its function is to map the received signal at time $t>0$, i.e., $\mathbf{y}_t^{\text{U}} \in \mathbb{C}^{M \times 1}$, to a sensing beam $\mathbf{s}_{t+1} \in \mathbb{C}^{M \times 1}$ at time $t+1$, which can be described by $g\left( \circ ;\boldsymbol{\vartheta }_t \right) :\mathbf{y}_t^{\text{U}} \mapsto \mathbf{s}_{t+1}$.
The UE NN module can also be described by a matrix product between two sub-modules, i.e., a complex NN mapping function and a wavenumber-domain mapping function, and can be described by
\begin{align}
	\mathbf{s}_{t+1}\left( \boldsymbol{\vartheta }_t \right) =\left(\mathbf{\Phi }_{\mathrm{U}}^{\left( \mathrm{e} \right)}\right)^*g^{\prime}\left( \mathbf{y}_t^{\text{U}};\boldsymbol{\vartheta }_t \right), \label{eq:ue_mapping}
\end{align}
where $g^{\prime}\left( \mathbf{s}_t; \boldsymbol{\vartheta }_t \right): \mathbb{C}^{M\times1} \mapsto \mathbb{C}^{| \mathcal{G} _{\boldsymbol{\kappa }}^{\left( \mathrm{e} \right)} | \times 1}$ denotes the complex NN mapping function, while $\mathbf{\Phi }_{\mathrm{U}}^{\left( \mathrm{e} \right)}: \mathbb{C}^{| \mathcal{G} _{\boldsymbol{\kappa }}^{\left( \mathrm{e} \right)} |\times 1}\longmapsto \mathbb{C}^{M \times 1}$ denotes the wavenumber-domain mapping function. 
To satisfy \eqref{c-2}, vector $\mathbf{s}_{t+1}$ is normalized according to $\mathbf{s}_{t+1}=\frac{1}{\sqrt{M}}\mathbf{s}_{t+1}\oslash \mathbf{s}_{t+1}^{\left| \cdot \right|}$.
The loss function on the UE side is given by
\begin{align}
	\mathcal{L} \left( \boldsymbol{\vartheta }_t \right) =\frac{1}{\sqrt{M}}\left| \left( \mathbf{s}_{t+1}^{}\left( \boldsymbol{\vartheta }_t \right) \oslash \mathbf{s}_{t+1}^{\left| \cdot \right|}\left( \boldsymbol{\vartheta }_t \right) \right) ^T\mathbf{y}_{t}^{\mathrm{U}} \right|, \label{loss-ue}
\end{align}
and the update rule is given by $\boldsymbol{\vartheta }_{t+1}=\boldsymbol{\vartheta }_t+w_{\mathrm{U}}\nabla _{\boldsymbol{\vartheta }_t}\mathcal{L} \left( \boldsymbol{\vartheta }_t \right) $, where $w_{\mathrm{U}}$ is the learning rate on the UE side.
Again, gradient ascent is utilized here.
The active-sensing-based method is illustrated by Fig. \ref{fig:NN}, and then summarized in \textbf{Algorithm 1}.

\begin{remark}
	\rm{The merits of the proposed active-sensing-based method can be summarized in what follows: 1)~{\textbf{Low overhead:}} The active-sensing-based method eliminates the need of codebooks and is carried out in a distributed manner;
	2)~\textbf{Fast convergence:}} By using $\mathbf{\Phi }_{\mathrm{B}}^{\left( \mathrm{e} \right)}$ and $\mathbf{\Phi }_{\mathrm{U}}^{\left( \mathrm{e} \right)}$, the beam alignment can be carrier out in a low-dimensional and diagonal domain.
	Therefore, the beam alignment process is accelerated;
	3)~{\textbf{Low complexity:}} $\mathbf{\Phi }_{\mathrm{B}}^{\left( \mathrm{e} \right)}$ and $\mathbf{\Phi }_{\mathrm{U}}^{\left( \mathrm{e} \right)}$ can also serve as two mapping functions, which maps from a low-dimensional vector space to a high-dimensional one, the number of trainable parameters is then reduced.
\end{remark}

\begin{algorithm}[hb] 
	\caption{Active-sensing-based algorithm to solve \eqref{object}.}
	\label{alg:narrow_active_sensing}
	\begin{algorithmic}[1] \small
		\STATE{\textbf{Initialization}: obtain $\mathbf{\Phi }_{\mathrm{B}}^{\left( \mathrm{e} \right)}$ and $\mathbf{\Phi }_{\mathrm{U}}^{\left( \mathrm{e} \right)}$ via estimating the positions of the UE/BS;
		initialize $\boldsymbol{\vartheta }_0$ and $\boldsymbol{\theta }_0$ and the maximum training round $T$;
		obtain $\mathbf{p}_1$ by feeding the BS with a zero vector.}
		\REPEAT
		\STATE{\textbf{UE}: a)~receive $\mathbf{y}_{t}^{\mathrm{U}}$ and obtain the sensing beam $\mathbf{s}_{t+1}$ by feeding $g\left( \circ ;\boldsymbol{\vartheta }_t \right)$ with $\mathbf{y}_t^{\text{U}}$; b)~obtain $\boldsymbol{\vartheta }_{t+1}$ by updating  $\boldsymbol{\vartheta }_{t}$ using \eqref{loss-ue}; c)~transmit $c_{\text{U}}$ using $\mathbf{s}_{t+1}$}
		\STATE{\textbf{BS}: a)~receive $\mathbf{y}_{t+1}^{\mathrm{B}}$ and obtain the probing beam $\mathbf{p}_{t+1}$ by feeding $f\left( \circ ;\boldsymbol{\theta }_t \right)$ with $\mathbf{y}_{t+1}^{\text{B}}$; b)~obtain $\boldsymbol{\theta }_{t+1}$ by updating  $\boldsymbol{\theta }_{t}$ using \eqref{loss-bs}; c)~transmit $c_{\text{B}}$ using $\mathbf{p}_{t+1}$.}
		\STATE{$t$ $\leftarrow$ $t+1$.}
		\UNTIL{$t$ reaches the maximum training round $T$.}
	\end{algorithmic}
\end{algorithm}

\section{Numerical Results} \label{sec:result}
In this section, the performance of the proposed method is evaluated by simulation results.
For the physical-layer parameters, the BS and the UE are equipped with $N=M=201$ half-wavelength antennas, the carrier frequency is set to $28$ GHz, the transmit powers are uniformly set to $P = P_{\text{B}}= P_{\text{U}} = 20$ dBm, the noise power is set to $-60$ dBm, and $\sigma_l^2$ is set to 0.01 for all $l$\cite{jiang2023near}.
For the neural network parameters, the layer structure of the BS NN module and that
of the UE NN module are given by $N \times 128 \times 64 \times |\mathcal{G} _{\mathbf{k}}^{\left( \mathrm{e} \right)} |$ and $M \times 128 \times 64 \times |\mathcal{G} _{\boldsymbol{\kappa}}^{\left( \mathrm{e} \right)} |$, respectively.
For activation functions, Linear is used for the last layer of the BS and the UE NN modules, and the ReLU is used for the rest. 
The learning rates are set to $w_{\text{B}} = w_{\text{U}} = 0.005$, and Adam is used as the optimizer for the modules. 
The results are averaged over $100$ repetitive experiments. 
The following three benchmarks are considered in our simulation:
\begin{itemize}
	\item {\textbf{Optimal with perfect CSI}}: By assuming the perfect channel information $\bf H$ is known, the optimal policy is obtained by the singular value decomposition (SVD) method, i.e., $\log _2( 1+P\left| \lambda _{\max}\{ \mathbf{H}^H\mathbf{H} \} \right|/\sigma ^2 )$, where $\lambda _{\max}{\{\mathbf{A} \}}$ extracts the most significant eigenvalue of ${\mathbf{A}}$.
	
	\item {\textbf{Ablation test}}: Matrices $\mathbf{\Phi }_{\mathrm{B}}^{\left( \mathrm{e} \right)}$ and $\mathbf{\Phi }_{\mathrm{U}}^{\left( \mathrm{e} \right)}$ in \eqref{eq:bs_mapping} and \eqref{eq:ue_mapping} are removed from the NN modules. 
	Therefore, $f^{\prime}\left( \mathbf{s};\boldsymbol{\theta } \right)$ and $g^{\prime}\left( \mathbf{p};\boldsymbol{\vartheta } \right)$ directly map the received signal at the BS/UE to the probing/probing beam. 

	\item {\textbf{Random policy}}: The BS and the UE select the probing and the sensing beam randomly. 
\end{itemize}

\begin{figure}[h]
	\centering
	\includegraphics[width=0.9\linewidth]{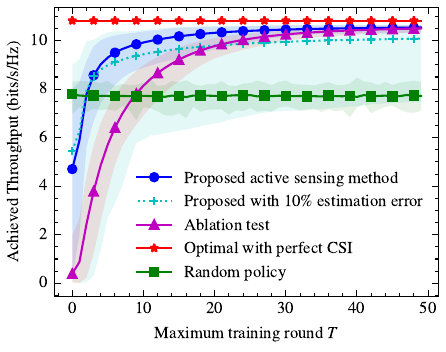}
	\caption{Throughput performance versus the maximum training round $T$, for $\theta = 90^\circ$, $d_{\text{BU}} = 15~{\rm m}$, and $L=3$.}
	\label{fig:throughput_T}
\end{figure}
In Fig. \ref{fig:throughput_T}, we evaluate the achieved throughput by varying the maximum training round $T$.
As $T$ increases, the achieved throughput after the beam alignment process increases and converges to the optimal results obtained by the optimal policy. 
It is noted that there is a gap between the proposed active-sensing-based method and the optimal policy.
The reason is two-fold: 1)~during the beam alignment process, the received signal at the BS/UE includes additive noise, which can lead to misalignment; 2)~the optimal policy via SVD is the theoretical upper bound under unit-power constraint and cannot be achieved under unit-modulus constraints\cite{heng2023grid}. 
Compared with the results of the ablation test, our proposed method can converge more quickly, which validates the effectiveness of $\mathbf{\Phi }_{\mathrm{B}}^{\left( \mathrm{e} \right)}$ and $\mathbf{\Phi }_{\mathrm{U}}^{\left( \mathrm{e} \right)}$.
Furthermore, with $\mathbf{\Phi }_{\mathrm{B}}^{\left( \mathrm{e} \right)}$ and $\mathbf{\Phi }_{\mathrm{U}}^{\left( \mathrm{e} \right)}$, the output layers of the two modules in the proposed method have lower dimension than that in the ablation test.
Therefore, the number of trainable parameters is decreased, leading to a fast convergence speed.    
Since the proposed algorithm needs the localization of the BS/UE to construct $\mathbf{\Phi }_{\mathrm{B}}^{\left( \mathrm{e} \right)}$ and $\mathbf{\Phi }_{\mathrm{U}}^{\left( \mathrm{e} \right)}$, we further investigate the impact caused by errors in localization estimation.
We can see from the results that our method can still converge quickly and achieve a good performance with a random estimation error ranging within $10 \%$ or $1.5~m$.
However, the error incurs a larger gap to the optimal policy.

\begin{figure}[h]
	\centering
	\includegraphics[width=0.95\linewidth]{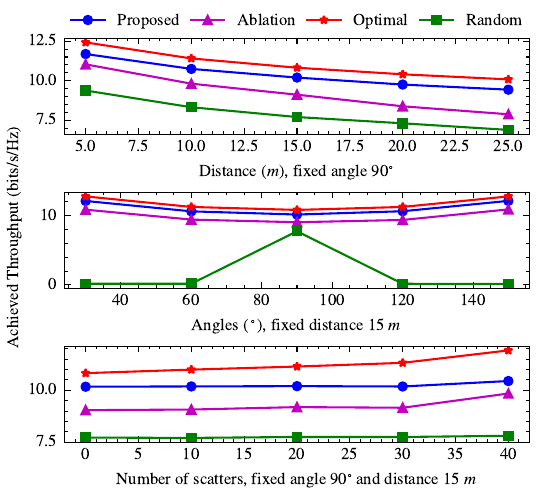}
	\caption{Throughput performance versus $d_{\text{BU}}$ (top), $\theta$ (middle), and $L$ (bottom), for $T=15$.}
	\label{fig:throughput_d_a_l}
\end{figure}

Fig. \ref{fig:throughput_d_a_l} shows the achieved throughput versus distances, angles, and number of scatters, respectively, for $T=15$.
Firstly, as $d_{\text {BU}}$ increases, the proposed method can achieve better performance than the ablation test, which is attributed to the fast convergence speed.
As $d_{\text{BU}}$ becomes larger, the gap between the optimal policy and the proposed method is narrowing.
The reason is that NFC channels can provide more degree-of-freedoms (DoFs) than the far-field case.
Therefore, the BS/UE needs more time to find the optimal beam when $d_{\text{BU}}$ is small.
Secondly, by varying $\theta \in [30^\circ, 150^\circ]$, the proposed method can achieve a stable near-optimal throughput performance.
It can also be observed that the random policy can only achieve decent performance when $\theta=90^\circ$.
Finally, as $L$ increases, the performance gap between our method and the optimal policy grows larger.
The reason lies in that by using $\mathbf{\Phi }_{\mathrm{B}}^{\left( \mathrm{e} \right)}$ and $\mathbf{\Phi }_{\mathrm{U}}^{\left( \mathrm{e} \right)}$, we extract LoS components of $\mathbf{H}$, while ignoring the rest NLoS components.
Hence, in a rich-scattering environment, our method ignores the benefits brought by these NLoS components.
However, in general cases, since LoS is dominant and scatterers are sparse, the proposed method can guarantee a good performance in practice.

\section{Conclusion} \label{sec:conclusion}
We proposed an active-sensing-based learning algorithm to solve beam alignment problem for NFC MIMO systems, which can alleviate overheads caused by beam sweeping.
Specifically, due to the dominance of LoS paths in mmWave communications, the truncated wavenumber-domain transform matrices were exploited to accelerate learning process.
The simulation showed that the proposed method can quickly converge and the effectiveness of wavenumber-domain transform matrices was validated by the ablation test.

\bibliographystyle{IEEEtran}
\bibliography{reference/mybib}

\end{document}